\documentclass[usenatbib,useAMS]{mn2e}
\usepackage{graphicx,portland,lscape}

\title[Galaxy counts and variances]{{\sc GalaxyCount}: a {\sc java} calculator of galaxy counts and variances in multiband wide-field surveys to 28 AB mag}

\author[Ellis \& Bland-Hawthorn]{S.C. Ellis\thanks{E-mail: sce@aao.gov.au} and J. Bland-Hawthorn\\ 
Anglo-Australian Observatory, P.O. Box 296, Epping, NSW 2121, Australia\\}

\date{Accepted...... Received .....}

\def\gtrsim{\mathrel{\hbox{\rlap{\hbox{\lower4pt\hbox{$\sim$}}}\hbox{$>$}}}}
\def\lsim{\mathrel{\hbox{\rlap{\hbox{\lower4pt\hbox{$\sim$}}}\hbox{$<$}}}}

\begin{document}
\maketitle

\begin{abstract}
We provide a consistent framework for estimating galaxy counts and
variances in wide-field images for a range of photometric bands.  The variances include both Poissonian noise and variations due to large scale structure.  We demonstrate that our statistical theory is consistent with the
counts in the deepest multiband surveys available.  The statistical
estimates depend on several observational parameters (e.g.\ seeing,
signal to noise ratio), and include a sophisticated treatment of detection completeness. The {\sc java} calculator is freely available$^{1}$ and offers the user the option to adopt our consistent
framework or a different scheme. We also provide a summary table of
statistical measures in the different bands for a range
of different fields of view.

Reliable estimation of the background counts has profound consequences
in many areas of observational astronomy.  We provide two such
examples.  One is from a recent  study of the Sculptor galaxy NGC\,300
 where stellar photometry has been used to demonstrate that the
outer disc extends to 10 effective radii, far beyond what was thought
possible for a normal low-luminosity spiral. We confirm this finding by a reanalysis of the background counts.  Secondly, we determine the luminosity function of the galaxy cluster Abell 2734, both through spectroscopically determined  cluster membership, and through statistical subtraction of the background galaxies using the calculator and offset fields.  We demonstrate very good agreement, suggesting that expensive spectroscopic follow-up, or off-source observations, may often be bypassed via determination of the galaxy background with {\sc GalaxyCount}.
\end{abstract}


\section{Introduction}

With the advent of wide-field imagers on high-quality observing 
sites, there are increasing numbers of very deep, wide-angle 
studies (e.g.\ nearby galaxies) that are subject to an unknown 
contribution of background galaxy counts. Even when the same
field has been observed in several photometric bands, it can be
very difficult to remove correctly the background galaxy counts
at a deep magnitude limit.

The issue of background contamination is likely to persist for many years.  Even with the fine resolution ($\approx0.1"$) of the Hubble Space Telescope (HST) the highest redshift sources are still largely unresolved (\citealt{ouc05b}).  Therefore star/galaxy discrimination of such objects awaits the development of multi-conjugate adaptive optics (MCAO) capable of spatial resolution on the order of tens of milli-arcseconds.  However, higher resolution observations will result in a smaller field of view (FOV), since the focal plane of any instrument can only be populated with a limited number of CCDs.   A ``wide-field imager'' in an era of MCAO is unlikely to exceed a few arcminutes (\citealt{bland06}).   A smaller FOV results in poorer statistical subtraction of background galaxies, due to larger relative uncertainty on the number counts. 

We provide a {\sc java} calculator\footnote{{\sc GalaxyCount} is freely available from http://www.aao.gov.au/astro/GalaxyCount}  for estimating these variations
in several photometric bands over a wide magnitude range. 
Our analytic approach is supported by the deepest surveys to date.
Our modelled magnitude limits are ($U,B,R,I,K$) $\approx$ (27, 29, 28, 28, 23) mag.

We first describe our model in Section~\ref{sec:model} and then present the results of the model with comparison to the Subaru Deep Field (SDF) in Section~\ref{sec:results}.   Section~\ref{sec:eg} describes worked examples of two practical applications of the model.  First to a study of star-counts in NGC\,300 by \citet{bland05}, and secondly to the determination of galaxy cluster luminosity functions.  Finally we discuss the results and present our conclusions in Section~\ref{sec:discuss}.

We have assumed a cosmology of $H_{0}=70$ km s$^{-1}$ Mpc$^{-1}$, $\Omega_{{\rm M}}=0.3$ and $\Omega_{\Lambda}=0.7$ throughout, although the results are not very sensitive to the exact choice of cosmology.

\section{Predicting counts and variances}
\label{sec:model}

The variance of galaxy number counts in a randomly placed cell of solid angle $\Omega$ is given by,
\begin{equation}
\label{eqn:var}
\sigma^{2}=n\Omega + n^{2}\int {\rm d}\Omega_{1} {\rm d}\Omega_{2} \omega(\theta_{12}), 
\end{equation}
where $n$ is the number density of galaxies, and ${\rm d}\Omega_{1}$ and  ${\rm d}\Omega_{2}$ are elements of the solid angle $\Omega$ separated by an angle $\theta_{12}$ (\citealt{pee80}).  The angular correlation function $\omega$ can be parametrized as,
\begin{equation}
\omega(\theta)=A\theta^{-\delta}.
\end{equation}

Following \citet{pee75} we change variables to rewrite equation~\ref{eqn:var} as,
\begin{equation}
\sigma^{2}=n\Omega + (n\Omega)^{2}A\theta_{\rm C}^{-\delta} \int \frac{{\rm d}^{2}x_{1} {\rm d}^{2}x_{2}}{|{\bf x_{1}}-{\bf x_{2}}|^{\delta}},
\end{equation}
where the integral on the right hand side is an integral over the cell scaled to unit area, and the scale factor of the cell $\theta_{\rm C}$ has been taken out of the integral.  For a square cell of side $\theta_{\rm C}$ the integral equates to 2.24 (\citealt{pee75}), for a circular cell the integral equates to $\approx 1.53$ with $\theta_{\rm C}$ equal to the diameter of the cell (found by numerical integration).  {\sc GalaxyCount} also allows oblong and elliptic window functions, which are integrated numerically within the code.

Thus in order to calculate the variance for any general observation of a given area and depth, it is necessary to know how the number density and the correlation function amplitude change as a function of magnitude in the appropriate waveband.  These will now be considered in turn.

\subsection{Number density}

A large number of published galaxy number count surveys exist in many different wavebands.  These have been compiled for the $UBRIK$ filters to yield the differential number density over a wide range of magnitudes ($18.375\le U\le 27.45$, $10.45\le B\le29.35$, $14.42 \le R \le 28.37$, $12.25\le I\le 28.25$, $9.37\le K\le 23.48$).  The list of sources used in this work is given in appendix~\ref{app:nsources}.

Thus the number density may be computed through integration over the required magnitude range.  In practice this is achieved by averaging all the data in bins of width 1 mag, and fitting a cubic spline through the resulting averages.  The spline is then integrated numerically.

\subsubsection{Completeness}

A very important caveat with the above method is that the published number density surveys have all been corrected for incompleteness at their faintest magnitudes in order to estimate the true number counts of galaxies (which is necessary to accurately derive cosmological models, etc.).  However, if the variance of galaxy counts is required to estimate how accurately the foreground/ background galaxies can be removed from an observation the above method will yield incorrect results.  The completeness of the observations must be accounted for.  Fewer counts will be made at the faintest magnitudes, and hence the contribution to the variance will decrease.  Because the number density is very high at faint magnitudes, the contribution from the faintest galaxies dominates the calculation, therefore it is imperative to accurately account for any incompleteness.

We define completeness as the number of galaxies detected divided by the true number of galaxies.  In practice the completeness is usually estimated by artificially generating galaxies of various magnitude and adding these to the images.  The completeness can then be estimated by running the source detection software and counting the number of artificial objects which have been recovered.  Thus the completeness will depend on the source detection algorithm and thresholds.  In particular the completeness function will change with the significance of the detection threshold, e.g.\ a 5$\sigma$ detection threshold (i.e.\ the flux of detected sources is $\ge 5\sigma$ above the background) will have brighter completeness limits than a 3$\sigma$ detection threshold.

The detection threshold also determines the effect of spurious detections, such that a higher threshold will have fewer spurious detections.  {\sc GalaxyCount} does not account for spurious detections, which would be few for typical detection thresholds of $\ge 3 \sigma$.  This should be borne in my mind by the user if it is expected that there are many spurious detections in their data.

The completeness for any observation depends on the telescope aperture, the combined throughput of the telescope, instrument and atmosphere, the exposure time, the seeing ($\Gamma$) and the signal to noise ratio (SNR) of the detections, providing that the seeing is well sampled.  Thus if the completeness function is known for any observation it may easily be scaled for any other observation, on the assumption that the shape of the curve does not change.   Furthermore the completeness function can be modelled using a sigmoid function of the form,

\begin{equation}
\label{eqn:sigmoid}
f=\frac{1}{1+e^{a(x-b)}},
\end{equation}
where $f=N_{{\rm measured}}/N_{{\rm true}}$ is the completeness, $b$ is the magnitude at which $f=0.5$, and $a$ (which has units of magnitudes$^{-1}$) determines the sharpness of the step.

\begin{figure}
\centering \includegraphics[angle=270,scale=0.33]{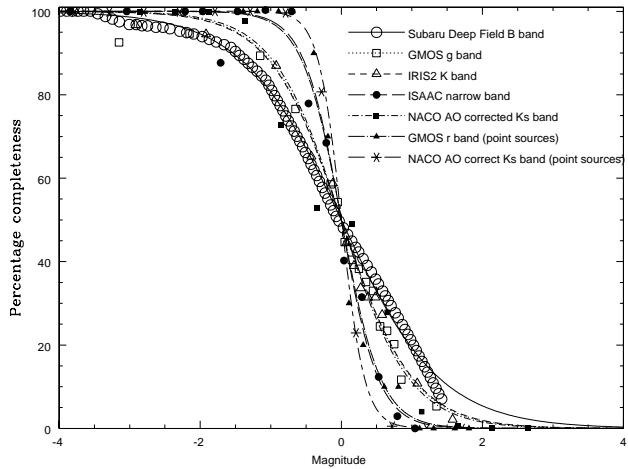}
\caption{Completeness functions compiled from the literature and our own data, all translated to have 50 per cent completeness at zero magnitude.  The points show the data from:  the Subaru Deep Field B band galaxy counts (open circles, \citealt{kas04}); the author's own GMOS g band observations of a redshift 0.7 cluster of galaxies (open squares); the author's own IRIS2 K band observations of the same field (open triangles); ISAAC 1.19$\mu m$ narrow band filter data of the Chandra Deep Field South (closed circles, \citealt{cub07}); NACO AO corrected observations (closed squares, \citealt{cre06}); the author's own GMOS r band imaging for point sources (closed triangles, \citealt{bland05}); NACO AO corrected observations for point sources (stars, \citealt{cre06}).  The lines show the best fitting sigmoid functions and range from $a=1.4$ for the SDF observations to $a=6.0$ for the NACO AO corrected point sources.}
\label{fig:sigmoids}
\end{figure}

\begin{figure}
\centering \includegraphics[scale=0.4]{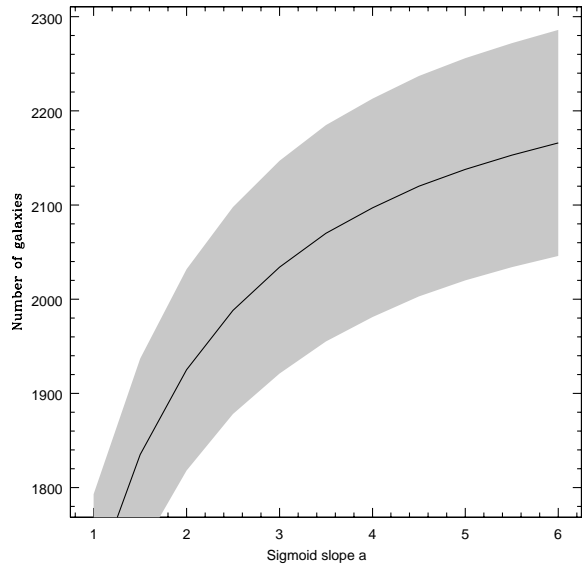}
\caption{The effect of the sigmoid slope on the resulting number counts and standard deviation for $18 \le B \le 27$ in a $5 \times 5'$ area, with a fixed 50 per cent completeness of 27 mag.  The solid line shows the number counts and the shaded area the 1$\sigma$ uncertainty on those counts.}
\label{fig:sigmoidslope}
\end{figure}

It is essential to use a sigmoid completeness function that correctly matches the observations in question, i.e. $a$ and $b$ must be correctly chosen.  This is very important because the galaxy number counts rise rapidly at faint magnitudes, and a change in the completeness function can have a large effect on the mean number counts and associated error.  In Figure~\ref{fig:sigmoids} we have compiled various completeness functions from the literature and our own data to show the spread in sigmoid slope which results from different object detection methods.  Object detection optimised for galaxies generally yield a flatter slope than those optimised for point source detection.  Similarly adaptive optics corrected images seem to result in a steeper slope than natural seeing images.  These trends are easily understood as extended sources of fixed magnitude will have a greater range of surface brightness than point sources of fixed magnitude.  Thus there is a larger spread in the sigmoid transition for  extended sources as some faint magnitude may have a high central surface brightness and thus be detected, whereas some bright magnitude sources may have an overall low surface brightness and thus be missed.   Only sigmoid functions derived from extended source detections should be used with {\sc GalaxyCount} because of this difference.  {\sc GalaxyCount} uses a default slope of $a=3.0$, which is typical for galaxy detections, however we stress that to obtain accurate results the completeness function for the users own data should be computed.  This illustrated in Figure~\ref{fig:sigmoidslope}, which shows the effect of varying the slope of the sigmoid function for B band observations with $18 \le B \le 27$ mag and a fixed 50 per cent completeness of B=27 mag.

{\sc GalaxyCount} uses a default sigmoid with $a=3.0$ which may  be used to provide an estimate the expected completeness of any observation by scaling $b$ appropriately.   The $b$ values were measured from the SDF $B$, $R$ and $I$ completeness functions in \citet{kas04}, and converted to Vega magnitudes using the transformations given in section~\ref{sec:mags} ($U$ was derived from the B band completeness function), and the SDF $K$ band completeness function was measured from \citet{mai01}.  In order to allow scaling from the SDF observations we have taken throughputs for Suprime-Cam $BRI$ (actually $i'$) from \citet{miy02} (convolved with Mauna Kea atmospheric transmission from \citealt{hoo04}), and for CISCO $K$ from \citet{mot02}, we have assumed the GMOS throughput for $U$.  We also offer the Gemini GMOS throughputs from \citet{hoo04} as a default option in the calculator (we assume that the $u'g'r'i'$ throughputs are appropriate for $UBRI$, and we assume the CISCO $K$ band throughput is appropriate for Gemini NIR imagers).

\begin{figure}
\centering \includegraphics[scale=0.4]{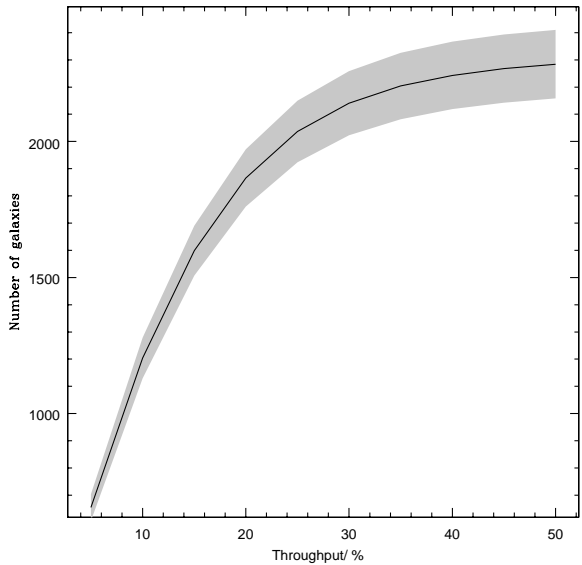}
\caption{The effect of the throughput on the resulting number counts and standard deviation for $18 \le B \le 27$ in a $5 \times 5'$ area, for a 595 minute exposure, with SNR=3, $\Gamma \approx 1"$, on an 8m aperture telescope.  The solid line shows the number counts and the shaded area the 1$\sigma$ uncertainty on those counts.}
\label{fig:through}
\end{figure}

Provision is made for the user to override any of our assumptions when using the calculator.  Observing conditions, telescope aperture and instrument throughput may all be freely specified, or a particular sigmoid completeness function may be input.  To give an impression of the importance of correctly specifying the appropriate observing conditions etc.\, Figure~\ref{fig:through} shows the effect of varying the throughput on the resulting $18 \le B \le 27$ mag number counts and standard deviation in a $5 \times 5'$ area, for a 595 minute exposure, with SNR=3, $\Gamma \approx 1"$, on an 8m aperture telescope.  Clearly it is important to use the correct input for the particular observations of concern.

\begin{figure}
\centering
\includegraphics[scale=0.42]{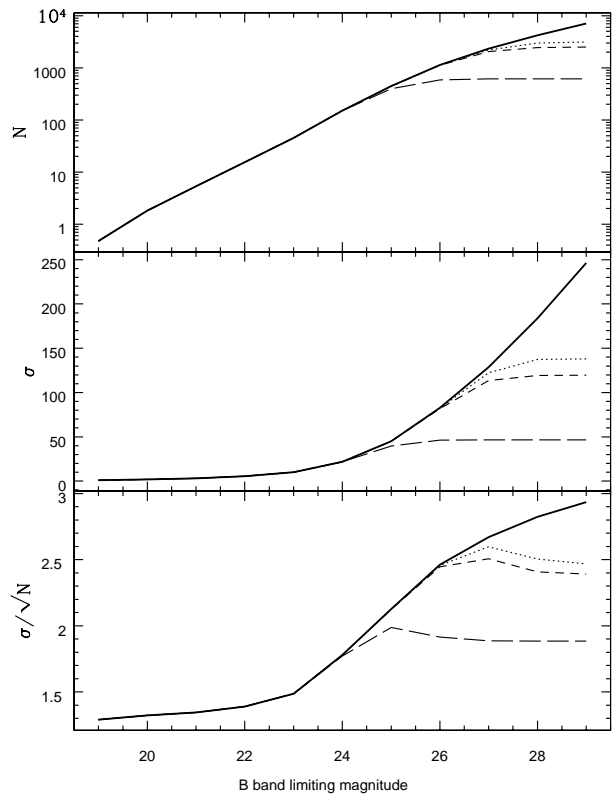}
\caption{The expected number, standard deviation, and standard deviation divided by square root of the number (i.e. the non-Poissonian component of the standard deviation) of galaxies in surveys of different magnitude limits, with a $5' \times 5'$ FOV.  The thick black line is for the total number of galaxies, the dotted and dashed lines include estimates for the incompleteness of the observations for exposure times of 595,  300 and 10 minutes, with SNR=3, $\Gamma \approx 1"$, with an 8m aperture telescope with 36\% throughput and deviate from the thick black line in that order.}
\label{fig:incomp}
\end{figure}

The difference in standard deviation resulting from incomplete and corrected observations is shown in Figure~\ref{fig:incomp}, for a $5'\times 5'$ FOV for a series of different exposure times and faint magnitude limits, with the other observing conditions the same as for the SDF.

Of course the number of galaxies in the field may already be known, and therefore {\sc GalaxyCount} includes provision for the user to enter his or her own value, as well as providing estimates from including and excluding incompleteness.
The standard deviation $\sigma_{2}$ associated with a measured number of galaxies, $N_{2}$, counted in a particular magnitude range and area, may be scaled from the predicted number and standard deviation, $N_{1}$ and $\sigma_{1}$, assuming the amplitude of the angular correlation function is the same in both observations, through the equation,

\begin{equation}
\label{eqn:scale}
\sigma_{2}=\sqrt{f^{2}\sigma_{1}^{2}+N_{1}f(1-f)},
\end{equation} 
which may be combined with equation~\ref{eqn:sigmoid}, to scale the results presented in Table~\ref{tab:main} for different values of completeness.

\subsection{Stellar contamination}

\begin{figure}
\centering \includegraphics[scale=0.4]{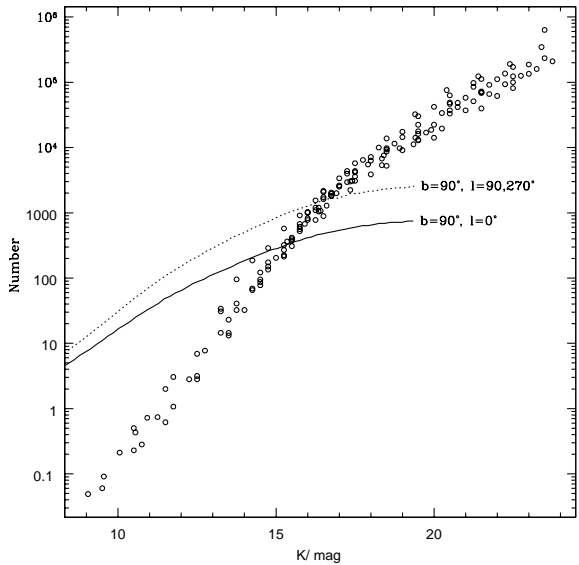}
\caption{The modelled star counts from \protect\citet{rob03} compared to galaxy counts compiled from the literature shown by the points.}
\label{fig:stars}
\end{figure}

The severity of stellar contamination will depend on the direction of the observations.  Stars are not expected to contribute significantly to the number counts for observations away from the Galactic plane (\citealt{rob03}).  Bright stars may be easily removed from the catalogues via visual inspection, radial profile fitting or automatic discrimination (e.g.\ \citealt{ber96}).  At fainter magnitudes the stellar contamination will not be significant.  Figure~\ref{fig:stars} compares the expected contribution of star counts at a Galactic latitude of $b=90$ degrees and longitudes $l=0,90,270$ degrees to the $K$ band galaxy counts compiled from the literature (see appendix~\ref{app:nsources}. If stellar contamination is expected to be a problem then it can be accounted for using synthetic models such as those of \citet{rob03}\footnote{http://bison.obs-besancon.fr/modele/}.

\subsection{Amplitude of the angular correlation function}
\label{sec:awmodel}

The amplitude of the angular correlation function decreases at fainter magnitudes, since the average redshift of the sample is larger and hence the volume being surveyed is larger, and thus the distribution of galaxies is approaching homogeneity.  

The functional dependence of $A$ on magnitude has been reported to be monotonically declining (e.g.\ \citealt{mcc01}; \citealt{wil03}; \citealt{coil04}) or to flatten off at the faintest magnitudes (e.g.\ \citealt{bra98}; \citealt{pos98}).  Both possibilities have been modelled in {\sc GalaxyCount}, and the user may choose between them.  The first case was simply modelled by a least squares fit to log$A$ vs. magnitude for data compiled from the literature.  However, to allow the models to be extrapolated to very faint magnitudes the second case has also been modelled, which allows for a flattening of $A$ at faint magnitudes.  The sources for the measurements of the angular correlation function used in this paper are listed in appendix~\ref{app:a}.

The amplitude of the angular correlation function has been modelled using theoretical arguments describing the evolution of clustering, in particular \citet{efs91} and \citet{pee80}.   The method takes a parametrized evolution of the two-point spatial correlation function and converts it to the expected dependence of $A$ on magnitude.  The two-point spatial correlation function is usually parametrized as,

\begin{equation}
\xi(r)=\left(\frac{r_{0}}{r}\right)^{\gamma}.
\end{equation}
The scaling of $\xi(r)$ with redshift, $z$, may be parametrized as (\citealt{gro77}),
\begin{equation}
\xi(r,z)=h(z)\left(\frac{r_{0}}{r}\right)^{\gamma},
\end{equation}
where
\begin{equation}
h(z)=(1+z)^{-(3+\epsilon)},
\end{equation}
where $r$ is the proper distance and $\epsilon$ describes the evolution of the clustering, with
\begin{equation}
\epsilon=\left\{ \begin{array}{ll}
        0 & \mbox{-- clustering fixed in proper co-ordinates} \\
        -1.2 & \mbox{-- clustering fixed in comoving co-ordinates}\\
        0.8 & \mbox{-- prediction of linear theory.}
        \end{array} \right.
\end{equation}

In all cases with $\epsilon>-1.2$ the comoving galaxy correlation length decreases monotonically with $z$.  {\sc GalaxyCount} uses $\epsilon=-1.2,0,0.8,3$ as selected by the user; $\epsilon=3$ was included to model the strong evolution of clustering observed at faint magnitudes (\citealt{coil04}).

For small angles $\omega(\theta)$ may be related to $\xi(r)$ as (\citealt{efs91}),

\begin{equation}
\label{eqn:awmodel}
\omega(\theta)=\sqrt{\pi}\frac{\Gamma[(\gamma-1)/2]}{\Gamma(\gamma/2)}\frac{A'}{\theta^{\gamma-1}}r_{0}^{\gamma},
\end{equation}
where $\Gamma$ is the complete gamma function and $A'$ is,
\begin{equation}
A'=\frac{\int_{0}^{\infty}g(z)\left(\frac{{\rm d}N}{{\rm d}z}\right)^{2}{\rm d}z}{\left[\int_{0}^{\infty}\left(\frac{{\rm d}N}{{\rm d}z}\right){\rm d}z\right]^{2}}
\end{equation}
and,
\begin{equation}
g(z)=\frac{h(z)}{d_{\rm A}^{\gamma-1}(z)({\rm d}r(z)/{\rm d}z)},
\end{equation}
where $d_{\rm A}$ is the angular diameter distance.

Thus, $A$ may be computed from equation~\ref{eqn:awmodel} by assuming a particular cosmology and adopting a form for ${\rm d}N/{\rm d}z$.  We have adopted an empirical relation for ${\rm d}N/{\rm d}z$ as a function of magnitude given by \citet{coil04},
\begin{equation}
\frac{{\rm d}N}{{\rm d}z}=z^{2}{\rm e}^{-z/z_{0}},
\end{equation}
where $z_{0}=-0.84+0.05I$.  Other bands were transformed to $I$ using the $z=0$ colours from the synthetic stellar population of \citet{bru03}, assuming an instantaneous burst of star-formation at $z=6.4$ and solar metallicity.

Note that the empirical form used for ${\rm d}N/{\rm d}z$ as a function of magnitude is not applicable for bright magnitudes.  Therefore we use a linear relation for magnitudes brighter than the point at which the slope of the theoretical curve and the best fitting linear relation are equal.  Although aesthetically more pleasing this in fact makes little difference to the results which are dominated by the much higher number counts at faint magnitudes.  The models are compared to observations (listed in appendix~\ref{app:a}) in Figure~\ref{fig:angcorrmodel}.  In most cases the monotonically decreasing model seems to be the best fit to the data.  

\begin{figure*}
\centering \includegraphics[scale=0.83]{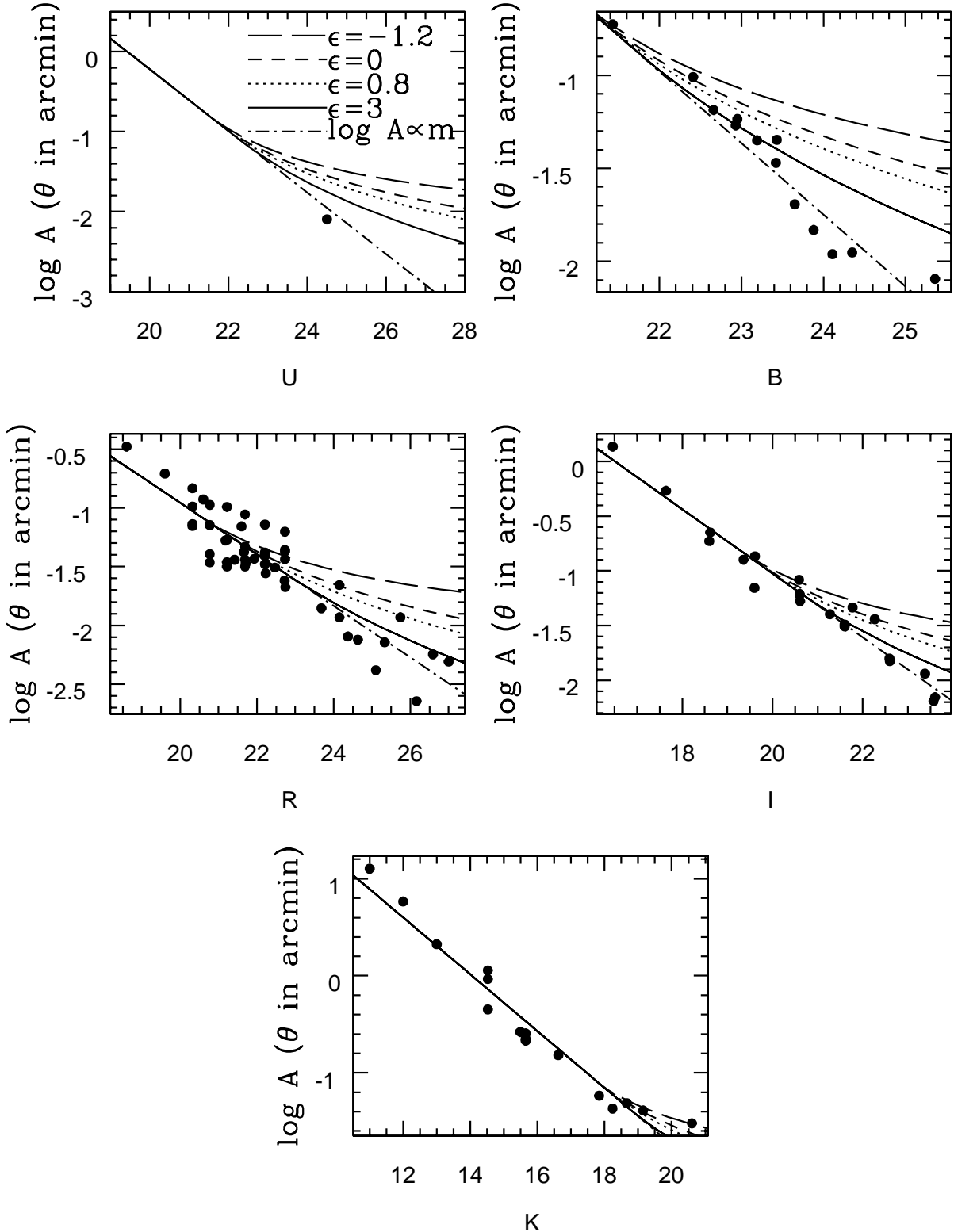}
\caption{The amplitude of the angular correlation function as a function of median apparent magnitude, compiled from the sources listed in appendix~\ref{app:a}.  The models described in section~\ref{sec:awmodel} are shown.  In most cases a linear model fits the data very well.}
\label{fig:angcorrmodel}
\end{figure*}

A subtle issue is that the fits to $A$ as a function of magnitude are calculated for a median magnitude.  In the sources that use limiting magnitudes we have converted to median magnitudes using our compiled number  counts within the magnitude limits.  However, $A$ is required over a range of magnitudes for use in equation~\ref{eqn:var}.  Therefore to calculate $\sigma^{2}$ between magnitude limits we substitute $n$ per magnitude for $n$ in equation~\ref{eqn:var}, and then integrate over magnitude between the limits in question, using the appropriate value for $A$ at each point.

\begin{figure}
\centering \includegraphics[scale=0.42]{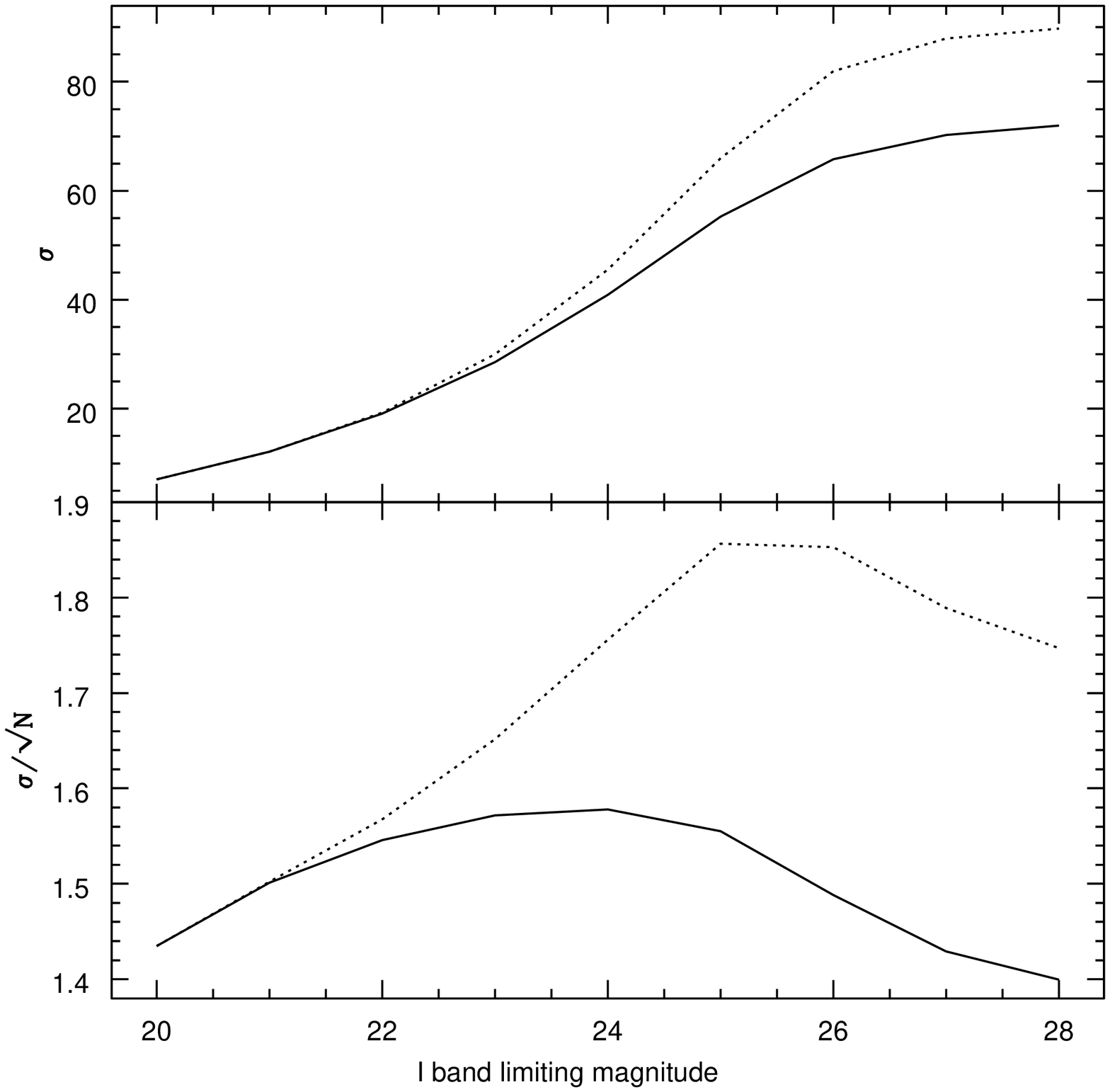}
\caption{The effect of the choice of model for $A$.  The solid line shows a model for a linear fit to log$A$ vs. magnitude; the dotted line shows a model with $\epsilon=3$, as described in the text.  The effect of the flattening of $A$ at faint magnitudes is to increase $\sigma$.}
\label{fig:awmodel}
\end{figure}

The effect of the chosen model of $A$ is shown in Figure~\ref{fig:awmodel} for a linear fit and a model with $\epsilon=3$.  The calculations were performed in the $I$ band, with a fixed bright magnitude of $I=18$, a FOV of 25 sq. arcmin., and a fixed exposure time of 300 minutes.  The effect of the flattening of $A$ at faint magnitudes is to increase $\sigma$.

\subsection{Magnitude systems}
\label{sec:mags}

{\sc GalaxyCount} uses data compiled from a large number of heterogeneous sources, listed in appendix~\ref{app:nsources}, to calculate the number density over a large range of apparent magnitudes.  These different sources comprise various filter and magnitude systems, and have been transformed to a common system of  `Kron' total magnitudes (\citealt{kro80}) in the Vega system ($B$ is Johnson and $R$ and $I$ are Kron-Cousins, see \citealt{met01}).  This system was chosen as most of the data were already in such a system, or a similar system, and therefore required the least amount of transformation.  However, we recognise that this system may not be the choice for many future astronomical systems, given the widespread use today of SDSS $u'g'r'i'z'$ filters, Petrosian magnitudes (\citealt{pet76}) and AB magnitudes (\citealt{fuk96}).  Thus in order to increase the capability of {\sc GalaxyCount} we have incorporated transformations to the $u'g'r'i'z'$ system, and list below the transformations to this and the AB systems.  These transformations are only approximate, since galaxies display a large range of colours, surface brightness profiles, spectra, etc. and thus no single transformation is appropriate for all galaxies.  Of course the user may simply compute the transformations appropriate to his or her own data, and input the transformed magnitudes.

Conversions to SDSS $u'g'r'i'z'$ magnitudes were calculated using the $z=0$ colours derived from a model spectrum of a passively evolving, solar metallicity galaxy, formed at $z=6.4$, from the libraries of \citet{bru03}.  The $z=0$ colours are assumed for all galaxies, and are listed below,

\begin{eqnarray}
u'\approx U+0.1 \nonumber \\
g'\approx B-0.3 \nonumber \\
r'\approx R+0.13\\
i'\approx I+0.1 \nonumber \\
z'\approx I-0.2. \nonumber
\end{eqnarray}

For comparison to the SDF (see the next section) the following conversions to AB magnitudes were used,

\begin{eqnarray}
\label{eqn:ABmag}
U_{{\rm AB}}\approx U+1.0 \nonumber \\
 B_{{\rm AB}}\approx B-0.1 \nonumber \\
 R_{{\rm AB}}\approx R+0.2 \\
 I_{{\rm AB}}\approx I+0.5 \nonumber \\
 K_{{\rm AB}}\approx K+1.9. \nonumber
\end{eqnarray}

For comparison to the Hubble Deep Fields (sectio~\ref{sec:hdf}) we use the following conversions given by \citet{met01},

\begin{eqnarray}
 B\approx F450+0.1 \nonumber \\
 R\approx F606-0.1 \\
 I\approx F814.\nonumber 
 \end{eqnarray}

\subsection{The effect of the window function}

\begin{figure}
\centering \includegraphics[scale=0.4]{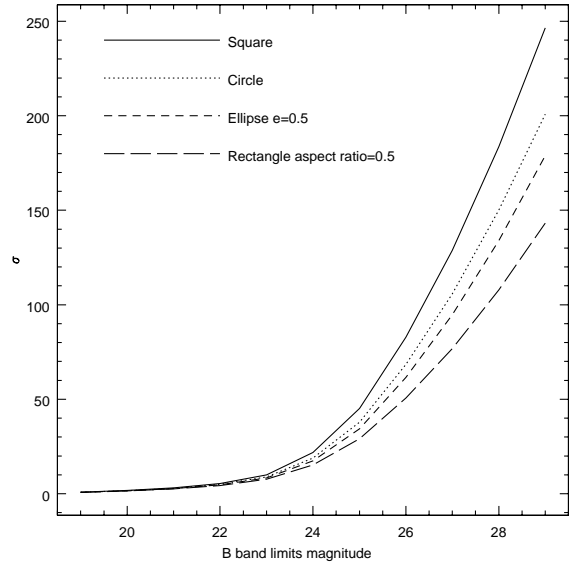}
\caption{The effect of the window function on the standard deviation for a constant area of 25 square arcminutes.}
\label{fig:window}
\end{figure}

{\sc GalaxyCount} allows a choice of window functions so different detector shapes and illumination patterns may be modelled.  The choices are square, circular, rectangular and elliptic.  Square and circular windows are treated separately from rectangular and elliptic window functions for the purpose of increasing the speed of the calculations.  In Figure~\ref{fig:window} we show the effect of the window function on the resulting galaxy counts and standard deviations for observations in the B band with a field of view of 25 square arcminutes in all cases.  The differences can be understood in terms of the different allowable separations between galaxies accommodated by the windows.

\section{Results}
\label{sec:results}

In this section we describe the general characteristics of the model via a comparison to data from the SDF.  
We demonstrate that the model reliably predicts accurate standard deviations for the associated number counts.  However the number counts are not always reliably predicted by the model, particularly in the $B$ band.  This is shown to be a consequence of the large discrepancy between published number densities at faint $B$ band magnitudes. 
 We also present general results of the model for different magnitude limits, filters, and areas.

Note well that the model works in the Vega magnitude system.  Therefore we have converted to the SDF AB magnitudes system using the transformations given in equation~\ref{eqn:ABmag}.

\subsection{Characteristics of the model and comparison to the Subaru Deep Fields}

Number counts and standard deviations have been computed from the SDF by random selection of square fields wholly contained within the SDF (and avoiding regions near bright stars and bad columns) and counting the number of galaxies.  Following \citet{kas04} we distinguish stars from galaxies using the following conditions:  if $mag \le 20$ and starclass$<0.99$, or if $20>mag\le 24$ and ln(isoarea $\times$ pix$) \ge (24-mag)/150$, or if $mag>24$, then we consider the object to be a galaxy, where starclass is the {\sc SExtractor} star/galaxy classifier, isoarea is the isophotal area above the {\sc SExtractor} analysis threshold and pix is the pixel scale.  For the discussion below 100 $2 \times2'$ square fields were randomly chosen in each band with a fixed bright magnitude limit of 20 mag in each band.  Integral counts are considered in all cases.

{\sc GalaxyCount} can be used to estimate the number counts of objects within the FOV and magnitude range, from which the standard deviation will be derived, or if the number counts of galaxies are already known they can be input directly into the calculation.  We present the results of three models with different methods of arriving at the number counts.  All models use $\epsilon=3$ for the variation of the amplitude of the correlation function which produces a higher standard deviation at faint magnitudes than a log($A) \propto m$ model, because $A$ becomes flatter at faint magnitudes for the $\epsilon$ models (see equation~\ref{eqn:var} and Figure~\ref{fig:awmodel}).

\begin{figure*}
\centering \includegraphics[scale=0.7]{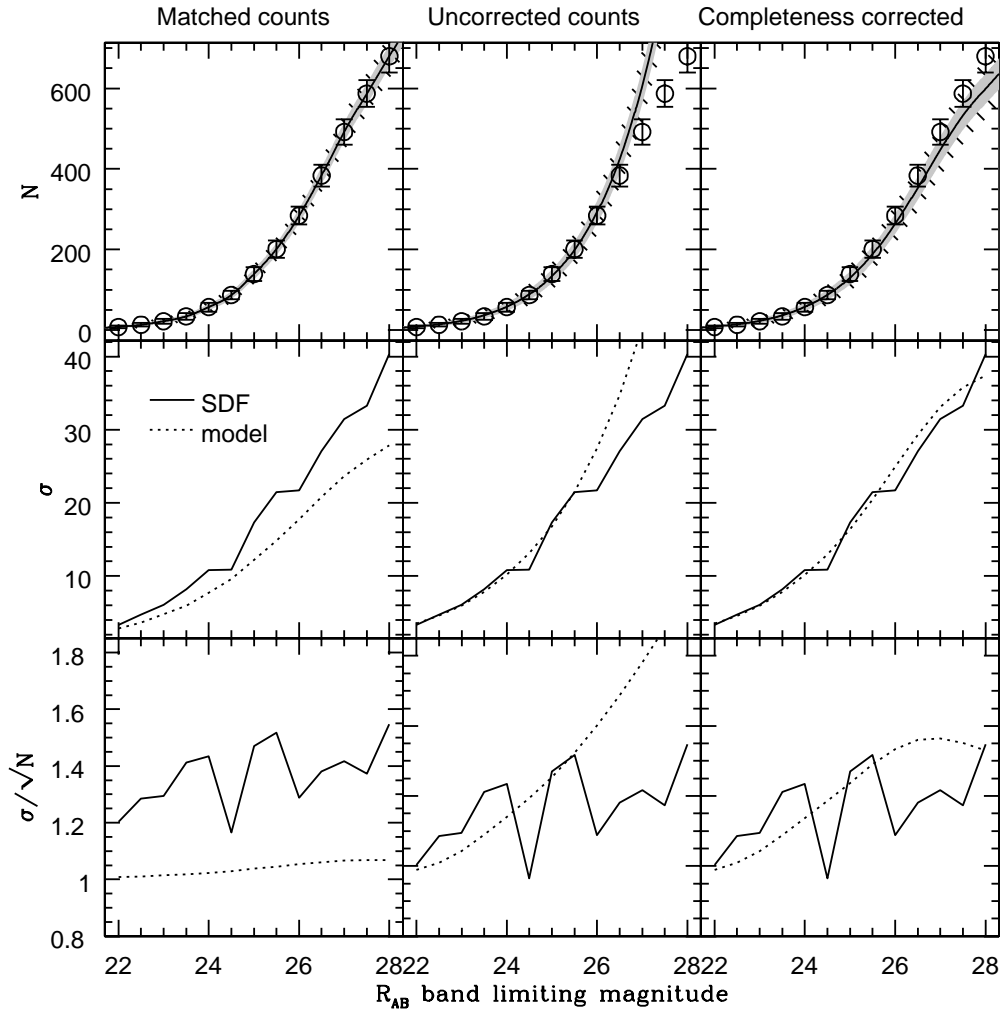}
\caption{Comparison of the model to the SDF $R$ band.  The model number counts have been calculated for an $\epsilon=3$ model.  The left hand panels show the data matched exactly to the SDF number counts.  The middle panels show a model with number counts calculated from number density surveys (see appendix~\ref{app:nsources}). The right hand panels show a model with number counts calculated from published surveys, but corrected for the incompleteness of the SDF observations.   The top panels show the SDF mean number counts and the associated stand deviation, as open points with error bars.  The grey area represents the predicted 1$\sigma$ uncertainty from the model, and the hatched area the 2$\sigma$ uncertainty.  The middle panels show the increase in $\sigma$ as a function of limiting magnitude, and  the bottom panels show the variation in $\sigma/\sqrt{N}$, i.e.\ the non-Poissonian component of the variance.}
\label{fig:rmodel}
\end{figure*}

The first model uses number counts taken directly from the SDF.  This allows us to examine the predicted standard deviation in isolation from the predicted number counts.   The results are shown in the left hand panels of Figure~\ref{fig:rmodel}.   The top panel shows the $R$ band SDF counts and the model.  The grey area shows the 1$\sigma$ uncertainty around the number counts, and the hatched area shows the 2$\sigma$ uncertainty.  The open points are the SDF mean number counts and the error bars are 1$\sigma$.    The middle panel shows the variation in the total uncertainty, $\sigma$ with limiting magnitude.  The bottom panel shows the variation in the non-Poissonian component of the uncertainty.  Whilst the model provides a reasonable match to the observed data for the total uncertainty, the non-Poissonian uncertainty is underestimated.  The reason for the disagreement is that although $n$ has been exactly specified, the variation of $n$ with magnitude has not and therefore an average value for $A$ must be used leading to the inaccuracy.

The second model has number counts calculated from published galaxy number density surveys.  The results are shown in the middle panels of Figure~\ref{fig:rmodel}, and the same general trends are seen as for the first model.  At the faintest magnitudes this model over predict the expected number of galaxies.  This is to be expected since the published number densities have been corrected for incompleteness, whereas our SDF measurements have not.  The over-prediction of the number counts contributes to  the standard deviation being over-estimated.

In the third model, shown in the right hand panels of Figure~\ref{fig:rmodel}, the incompleteness of the SDF observations has been taken into account to avoid the over-prediction of galaxy counts at faint magnitudes, using a sigmoid with a slope of $a=1.4$, which was found by fitting to the completeness functions from \citet{kas04}.  This model matches the data extremely well, both in terms of the total uncertianty and the non-Poissonian uncertainty.

The models were also tested for in the $B$ and $i'$ bands, and show very similar results.  However, for the number counts at faint limiting $B$ and $i'$ band magnitudes the model overcompensates for the incompleteness despite the fact that the incompleteness function is taken directly from the SDF observations of \citet{kas04}.  Consequently the predicted uncertainties also fall below the measured values.  

\begin{figure*}
\begin{minipage}[c]{0.4\textwidth}
\centering \includegraphics[scale=0.32]{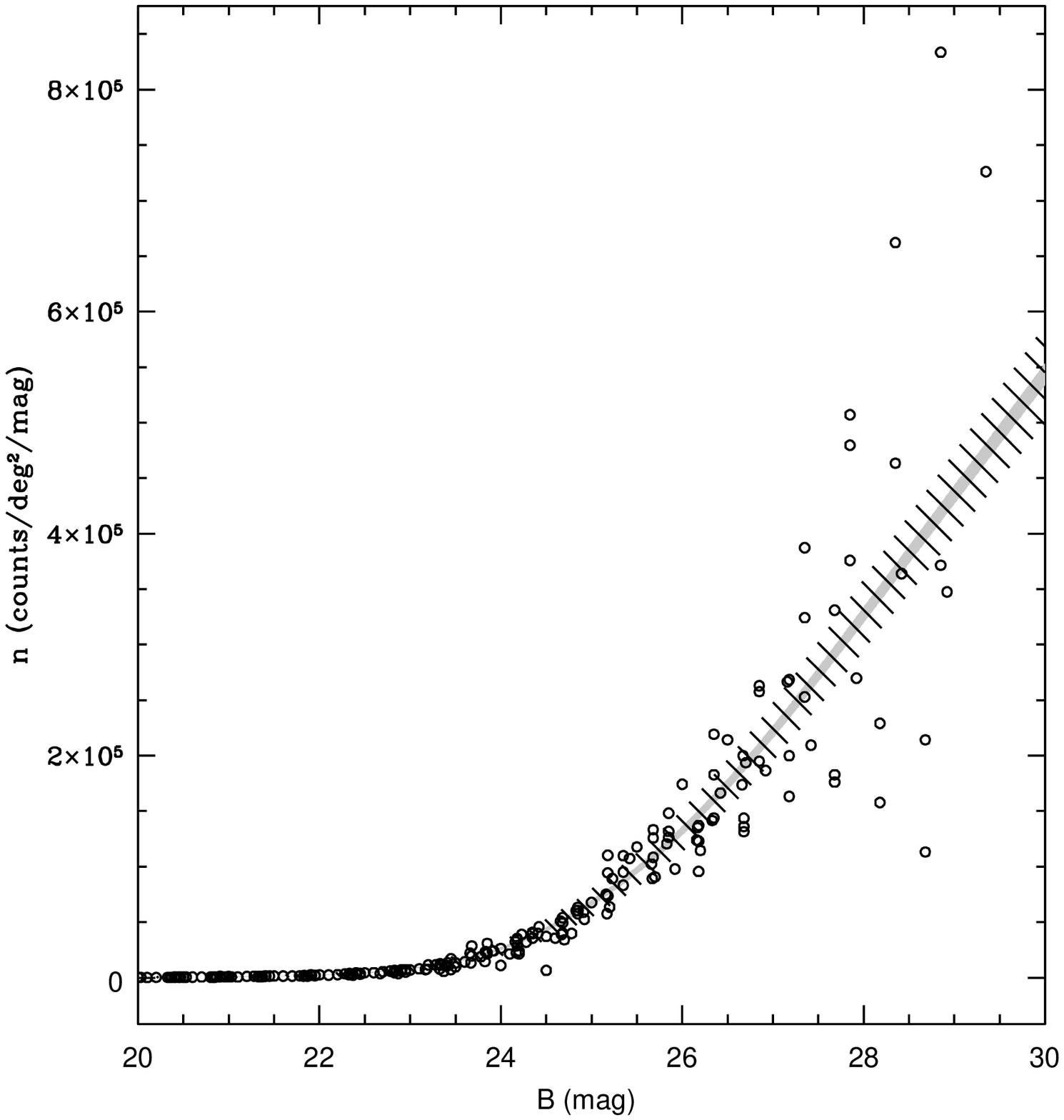}
\end{minipage}%
\begin{minipage}[c]{0.4\textwidth}
\centering \includegraphics[scale=0.32]{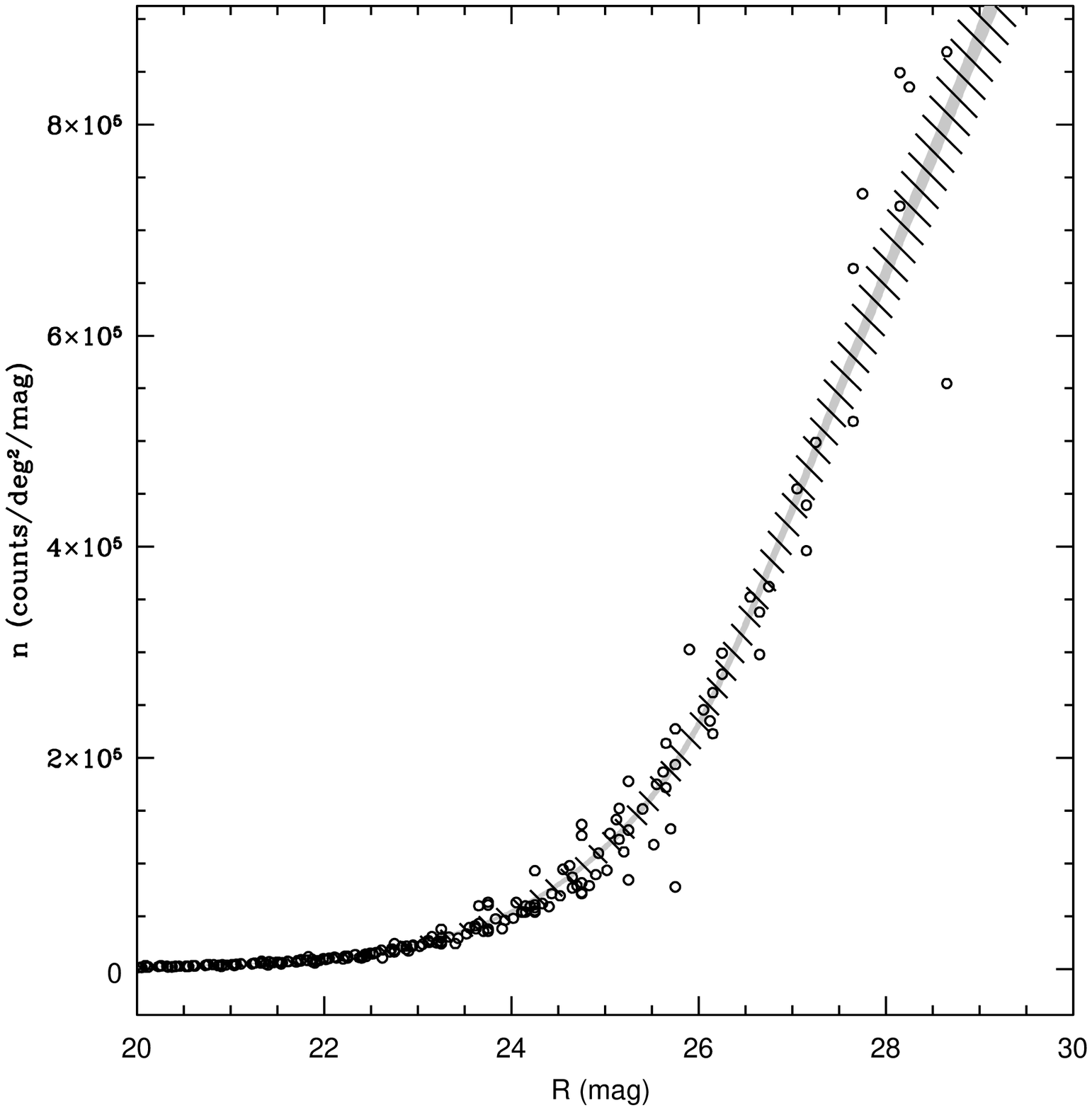}
\end{minipage}%
\caption{The number density of galaxies vs. magnitude, with $B$ on the left and $R$ on the right.  The grey shaded curves show the number density and 1$\sigma$ uncertainty expected from cosmic variance.  The hatched curves show the cosmic variance as measured in a survey with the the area of the Hubble Deep Field, and subsequently extrapolated to 1 deg$^{2}$.}
\label{fig:systematic}
\end{figure*}

These discrepancies  may be due in part to uncertainties in correcting for both the magnitude systems and the different filters used by the SDF and the collated number counts.  Furthermore, the number counts are calculated from a cubic spline fit to an average of compiled number density surveys from the literature.  The $B$ band SDF counts are systematically higher than the other surveys which reach faint magnitudes, and thus the average counts are already lower than the SDF counts before the incompleteness correction is applied.  However, this still leaves the problem that the uncertainties in number counts are under-predicted by the model.  We have shown in Figure ~\ref{fig:rmodel} that the model will predict the uncertainties very accurately if the number counts as a function of magnitude are correct.  Therefore it seems that the $B$ band SDF counts differ from the average counts by more than would be expected from cosmic variance.  This is not to say the $B$ band SDF counts are wrong, but rather that there are systematic differences between published number count surveys that are larger than would be expected from cosmic variance, most likely resulting from uncertainty in correcting for incompleteness and subsequent extrapolation to the standard one square degree for which number densities are quoted.  This is illustrated in Figure~\ref{fig:systematic}, which shows  number density vs. magnitude.  It is clear that the scatter dramatically increases at $B\gtrsim 25$, which is unlikely to be a real effect due to cosmic variance.  In comparison the scatter in the $R$ band counts is much more uniform at faint magnitudes.  The expected uncertainty due to cosmic variance is also shown in these plots.  The grey curves show the expected cosmic variance from a survey of 1 deg$^{2}$; the hatched curves show the expected cosmic variance from a survey of $1.17 \times 10^{-3}$ deg$^{2}$ (the area of the Hubble Deep Field) extrapolated to 1 deg$^{2}$.   Both simulations are assumed to be 100\% complete, an additional small correction would be needed for incomplete surveys.  The published number densities differ within cosmic variance at bright magnitudes but have a larger scatter at faint magnitudes.  Whilst this is markedly more pronounced in the $B$ band it is also true in the $R$ band.  The extra scatter at faint magnitudes can be attributed to systematic errors in accounting for the incompleteness of the surveys, or to systematic errors in consolidating the different magnitude systems of each survey.  Note that an accurate number density is crucial to correctly calculate the associated uncertainties, and thus the average presented here may fail to achieve this if systematic errors really are present.  We note though that the user is free to specify the number density exactly, and so may choose to rely on a particular survey of his or her own choice e.g. the Hubble Deep Fields, or to input a number density from his or her own survey.

\subsection{General results}

Table~\ref{tab:main} shows the results for a selection of magnitude ranges and areas.  These values were computed using the $\epsilon=3$ model, and assume 100\% completeness at all magnitudes.  The variances may be easily scaled to any number counts within the same magnitude range and area, assuming that the amplitude of the correlation function does not change, equation~\ref{eqn:scale}.  After appropriate scaling, equation~\ref{eqn:sigmoid} may be used to estimate the completeness if the number of galaxies is unknown, e.g.\ if star-galaxy discrimination is not possible.

\begin{table*}
\caption{Galaxy counts and standard deviations for a range of filters, magnitudes and areas.  These values were computed using the $\epsilon=3$ model and assume 100\% completeness at all magnitudes.  The standard deviations may be scaled for incompleteness or higher number counts using equation~\ref{eqn:scale}.}
\label{tab:main}
\begin{tabular}{ccccccccccccccc}
&\multicolumn{14}{c}{Counts and standard deviation within the given area} \\
&\multicolumn{2}{c}{$30''\times30''$}&\multicolumn{2}{c}{$1'\times1'$}&\multicolumn{2}{c}{$5'\times5'$}&\multicolumn{2}{c}{$10'\times10'$}&\multicolumn{2}{c}{$15'\times15'$}&\multicolumn{2}{c}{$30'\times30'$}&\multicolumn{2}{c}{$1^{\circ}\times1^{\circ}$}\\
Mag range & $N$ & $\sigma$ &$N$ & $\sigma$ &$N$ & $\sigma$ &$N$ & $\sigma$ &$N$ & $\sigma$ &$N$ & $\sigma$ &$N$ & $\sigma$ \\ \hline

$19\le U\le 21$ & 0  & 0    & 0 & 1       & 7 & 34        & 29 & 11               & 66 & 19              & 263 & 55                  & 1050 & 160 \\
$19\le U\le 23$ & 1 & 1     & 3 & 2       & 76 & 15    & 306 & 41          & 688 & 76               & 2750 & 220             & 11010 & 670 \\
$19\le U\le 25$ & 7 & 3    & 27 & 6      & 669 & 58 & 2680 & 160      & 6020 & 310         & 24080 & 920            & 96300 & 2700 \\
$19\le U\le 27$ & 30 & 6 & 121 & 15  & 3020 & 140 & 12100 & 410 & 27210 & 780     & 108900 & 2300       & 435400 & 7000 \\
$15\le B \le 17$ & 0 & 0   & 0 & 0         & 0 & 0        & 0 & 1                  & 0 & 1                     & 2 & 2                         & 6 & 6 \\
$15\le B \le 19$ & 0 & 0   & 0 & 0         & 1 & 1        & 3 & 3                  & 6 & 4                     & 24 & 12                       & 95 & 35 \\
$15\le B \le 21$ & 0 & 0   & 0 & 0         & 6 & 3        & 22 & 8                & 50 & 14                & 199 & 40                   & 800 & 120 \\
$15\le B \le 23$ & 0 & 1   & 2 & 1         & 45 & 10        & 181 & 26           & 408 & 47              & 1630 & 140                &  6530 & 400 \\
$15\le B \le 25$ & 4 & 2   & 18 & 5       & 449 & 45   & 1800 & 130         & 4040 & 240          &16180 & 710            &  64700 & 2100 \\
$15\le B \le 27$ & 23 & 6  & 93 & 13   & 2330 & 130 &  9310 & 370      & 20940 & 710       & 83800 & 2100         &  335100 & 6400 \\
$15\le B \le 29$ & 70 & 10  & 282 & 24 & 7050 & 250 & 28190 & 720  & 63400 & 1400       & 253700 & 4100        & 1015000 & 12000 \\
$15\le R\le 17 $ & 0 & 0    & 0 & 0        & 0 & 1        & 2 & 1                  & 4 & 2                      &14 & 5                        &  58 & 14 \\
$15\le R\le 19 $ & 0 & 0    & 0 & 0        & 4 & 2        & 18 & 6                & 40 & 10                   & 162 & 26                  &   646 & 75 \\
$15\le R\le 21 $ & 0 & 1    & 1 & 1        & 33 & 8      & 131 & 20           & 295 & 36              & 1179 & 100                &   4720 & 290 \\
$15\le R\le 23 $ & 2 & 1    & 7 & 3        & 184 & 22 &  735 & 60          & 1650 & 110             & 6610 & 320             &   26450 & 950 \\
$15\le R\le 25$  & 10 & 3  &  40 & 8    &  992 & 67 &  3970 & 190    &  8930 & 360          & 35700 & 1100           &   142900 & 3200 \\
$15\le R\le 27$  & 44 & 8  &  175 & 19 & 4380 & 190 &17510 & 540 & 39390 & 1000       & 157600 & 3100     & 630300 & 9300 \\
$15\le R\le 28$  & 83 & 11 & 331 & 28 & 8260 & 290 & 33060 & 840 & 74400 & 1600  & 297500 & 4800      & 1190000 & 15000 \\
$14\le I \le 16 $  & 0 & 0      & 0 & 0        & 0 & 0       & 1 & 1                  & 2 & 2                     & 7 & 4                        & 28 & 11 \\
$14\le I \le 18 $  & 0 & 0      & 0 & 0        & 4 & 2       & 14 & 6                & 32 & 11                 & 129 & 31                  &  514 & 89 \\
$14\le I \le 20 $  & 0 & 1      & 1 & 1        & 28 & 8     & 110 & 21           & 248 & 37              & 990 & 110                  &  3970 & 320 \\
$14\le I \le 22 $  & 2 & 1      & 6 & 3        & 157 & 21 & 627 & 56         & 1410 & 100            & 5640 & 300             &  22560 & 900 \\
$14\le I \le 24 $  & 7 & 3      & 28 & 6      & 712 & 52 & 2850 & 140     &  6410 & 270         & 25630 & 790           &  102500 & 2400 \\
$14\le I \le 26 $  & 32 & 7    & 128 & 15 & 3190 & 140 & 12760 & 420 & 28710 & 790     & 114800 & 2400      & 459300 & 7100 \\
$14\le I \le 28  $ & 113 & 13 & 451 & 33 & 11280 & 340 & 45110 & 1010 & 101500 & 1900 & 406000 & 5700 & 1624000 & 17000 \\
$13\le K \le 15$  &  0 & 0      & 0 & 0       & 1 & 1       & 5 & 3                  & 11 & 5                   & 42 & 14                   &  169 & 39 \\
$13\le K \le 17$  &  0 & 0      & 1 & 1       & 16 & 6     & 64 & 15             & 144 & 28              & 577 & 79                 &   2310 & 230 \\
$13\le K \le 19 $ &  1 & 1      & 4 & 2      & 110 & 18 &  441 & 48          &  993 & 87             & 3970 & 260            &   15880 & 760 \\
$13\le K \le  21$ &  5 & 3      & 22 & 5    & 543 & 44 &  2170 & 120       &  4890& 230         & 19560 & 670         &   78200 & 2000 \\
$13\le K \le  23$ &  20 & 5    & 81 & 11 & 2022 & 96 & 8090 & 270   &  18200 & 510        & 72800 & 1500        &   291200 & 4500 \\

\end{tabular}
\end{table*}

We have explicitly demonstrated the reliability of the $B$, $R$ and $I$ band predictions above, via comparison with the SDF.  However, the $U$ and $K$ bands have not been checked since there are no sufficiently large and deep surveys to perform a satisfactory random counts-in-cells test.  The $K$ band has good statistics on the number counts and the angular-correlation function, so there is no reason to suspect that the results will be very wrong but until they are checked against observations we cannot know for certain.  The $U$ band is still more uncertain since the variation of $A$ with magnitude is unknown, and so the $B$ band function has been assumed, and normalised to the $U$ band.

\section{Worked examples}
\label{sec:eg}

\subsection{NGC\,300}
\label{sec:ngc300}

We now provide a worked example to emphasise the importance of
accurately determining the background galaxy counts in a given
photometric band. \citet{bland05} obtained deep
observations of the stellar population in the outer disc of NGC\,300, a late-type spiral in the Sculptor group.  They used the Gemini
Multi-object Spectrograph (GMOS) on the Gemini South 8m telescope
in exceptional conditions (0.6" FWHM seeing).  At a 3$\sigma$ point source
detection threshold of $r' = 27.0$ mag, they were able to
trace the stellar disc out to a radius of 24', or $2.2\times R_{25}$ where
$R_{25}$ is the 25 mag arcsec$^{-2}$ isophotal radius.  This corresponds
to about 10 scale lengths in this low-luminosity spiral ($M_{B}=
-18.6$), or about 14.4 kpc at a distance of 2.0 Mpc.

These authors, and others (e.g.\ \citealt{fer05}; \citealt{irw05}),
demonstrate the profound importance of using star counts for tracing
the outer reaches of galaxies. \citet{bland05} reach
an effective surface brightness of 30.5 mag arcsec$^{-2}$ (2$\sigma$) at
55\% point source completeness which doubles the known radial extent of the optical
disc.  These levels are exceedingly faint in the sense that the
equivalent surface brightness in $B$ or $V$ is about 32 mag arcsec$^{-2}$.

In the NGC\,300 study, the authors found no evidence for truncation
of the stellar disc such that the disc may extend even further. \emph{But
this conclusion is sensitive to the assumed background galaxy counts}.
The disc can be forced to truncate at smaller radius if the background
galaxy counts are close to or exceed 200 gals arcmin$^{-2}$. However,
the authors found in favour of a lower background (130 gals arcmin$^{-2}$) by assuming that the outermost field is dominated by
background galaxies.

Note that in order to statistically subtract background galaxies it is often necessary to observe ``blank-sky'' fields, for an equal amount of time as the field of interest.  The {\sc java} calculator obviates this expensive requirement (which in practice is not often met).  Provided that the completeness of the target observations can be calculated, the deep galaxy surveys incorporated into the calculator can be used to subtract the contaminating galaxies and to estimate the associated error.

We now use {\sc GalaxyCount} to predict the expected galaxy counts
in $r'$ for a 5.5'$\times$5.5' field. This takes into account
the seeing (0.6"), SNR=3$\sigma$, the telescope aperture (8m), total system throughput=39\%,
and the exposure time (8100s). For our values (in this case we use $R=r'-0.2 < 26.8$, which is appropriate for stellar populations in the outer discs of galaxies, see \citealt{bland05}), we determine 
85 $\pm 11$ galaxies arcmin$^{-2}$, or 2560 $\pm$ 110 galaxies over the GMOS FOV, where the errors are 1$\sigma$, calculated using an $\epsilon=3$ model.  This is equivalent to a total completeness of $\approx 56\%$   for $R\le26.8$ for galaxies as predicted by the calculator.

We now consider the impact of the background galaxy counts on the
luminosity profile derived from the stellar counts.  The primary aim of \citet{bland05} was to obtain deep observations of the outer disc of NGC\,300, and hence only a section of the disc was studied, located approximately 9kpc SW of the galaxy nucleus on the semi-major axis.  Star counts were made in arcs of fixed width centred on the galaxy nucleus, thus the area of each segment is different.  Consequently the uncertainty on the star counts and on the background galaxy counts will change as a function of radius. 



\begin{figure}
\centering \includegraphics[scale=0.4, angle=0]{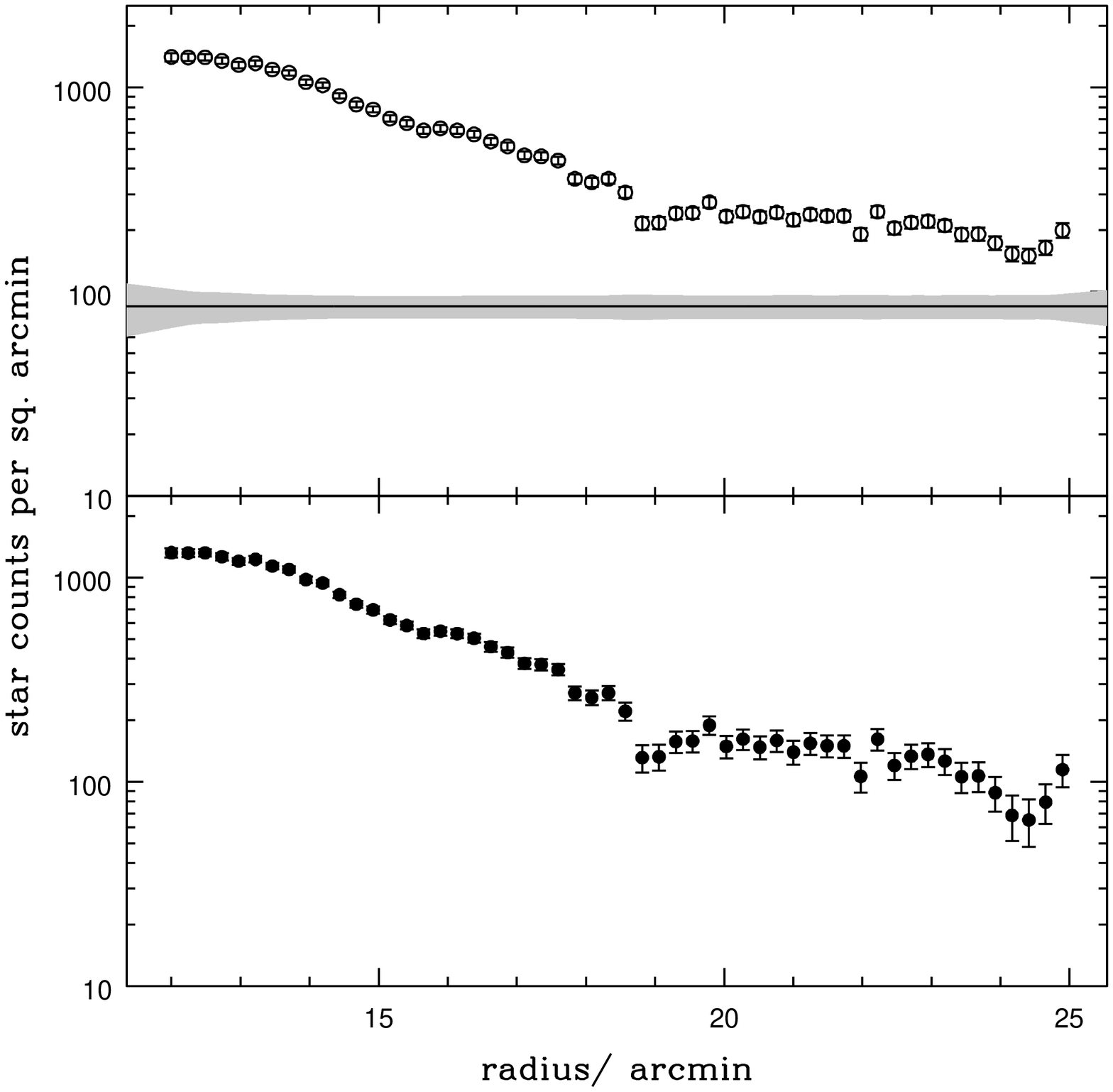}
\caption{The luminosity profile of NGC\,300.  The top panel shows the raw counts as the open symbols, with the associated error (calculated as the square root of the counts), the solid line represents the expected contribution to the raw counts from background galaxies, and the shaded region represents the uncertainty on this number as predicted by {\sc GalaxyCount}.  In the bottom panel the closed symbols represent the corrected star counts, and the combined errors.}
\label{fig:arcs}
\end{figure}

Figure~\ref{fig:arcs} reproduces the luminosity profile of NGC\,300 star counts.  There is no evidence for a truncation of the stellar disc; the star counts decline exponentially to the limits of the data, which is equal to about 10 scale lengths.  This confirms the findings of \citet{bland05}, but with the added assurance that the errors introduced through the subtraction of an estimated or a  theoretical number of background galaxies do not significantly affect the conclusions.  \citet{bland05} found they could generate a truncation of the disc if the background contains $\approx 220$ galaxies arcmin$^{-2}$, but this is ruled out at high significance.

\subsection{Galaxy cluster luminosity functions}
\label{sec:lf}

Another application of {\sc GalaxyCount} is the statistical removal of foreground and background galaxies from clusters in order to compute the properties of the cluster members.  Here we provide an example of such use in the computation of the $b_{{\rm j}}$ band luminosity function (LF) of cluster Abell 2734.  This cluster was selected from the two degree field galaxy redshift survey (2dFGRS) study of  the cluster LF (\citealt{dep03}), based on its high membership of 127 galaxies with the selection of  \citet{dep03}.  The cluster has a redshift of $cz=18646$ km s$^{-1}$ and velocity dispersion of $\sigma=1038$ km s$^{-1}$. 

We have calculated the LF in three ways.  First we select all galaxies within a box of 4Mpc centred on the cluster having redshifts within $\pm 3 \sigma$ of the velocity dispersion.  This yields an accurate catalogue of cluster members from which the cluster LF is computed.  Secondly we count all galaxies within the 4Mpc box regardless of their redshifts, as if the redshifts were unknown.  We then use {\sc GalaxyCount} to estimate the number of interlopers as a function of magnitude and compute the LF based on these corrections.  Finally we select offset fields around the cluster from which to measure the background galaxy counts to compare with the standard practice of statistical subtraction of foreground and background galaxies.  

\begin{figure}
\centering \includegraphics[scale=0.4]{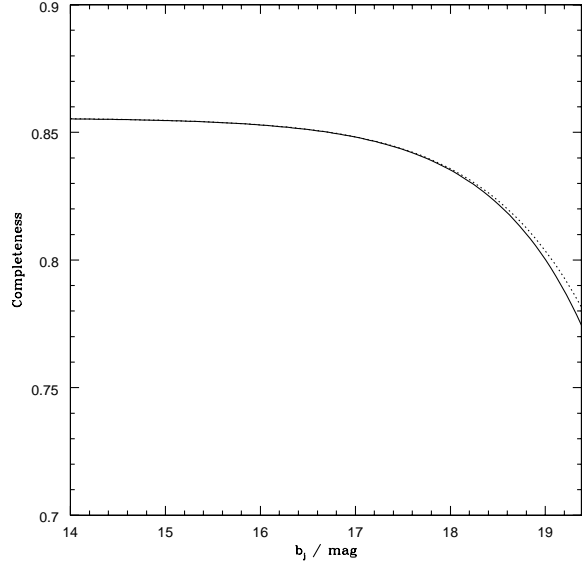}
\caption{The 2dFGRS completeness as a function of magnitude at the position of A2734 is shown by the solid line.  The dotted line is our sigmoid representation.}
\label{fig:2dfcompl}
\end{figure}

In order to use {\sc GalaxyCount} to model the 2dFGRS counts the completeness limit must be accurately modelled.  The 2dFGRS has a magnitude limit of $b_{\rm j}\approx19.45$ mag, but varies slightly as a function of position.  There is also a redshift completeness that is a function of magnitude.  \citet{col01} provide a model for the completeness function, and software is available from the 2dFGRS web pages\footnote{http://www.mso.anu.edu.au/2dFGRS/} to compute the necessary parameters to determine the final completeness as a function of magnitude.  We match our sigmoid model to their completeness function, as shown in Figure~\ref{fig:2dfcompl}.  Finally we convert $B=b_{\rm j}+0.2$. It is worth re-emphasising that accurate completeness functions are crucial for the success of modelling expected counts with {\sc GalaxyCount} since the number counts at faint magnitudes are so dominant.  Note though that accurate completeness functions can be readily computed from the imaging data in question.

The offset fields were chosen at 1 and 2 degree displacements in RA and dec around cluster centre giving 24 fields in total.  Computing the LFs for a large number of offset fields allows us to determine the relative merits of using {\sc GalaxyCount} against using offset fields, since different offset fields can give very different results.  Note that there are two methods in which the offset LFs may be computed: by using the total raw counts in the cluster and offset fields, or by counting only those galaxies for which redshifts are known.  The former method mimics how the technique would usually be employed in standard observational practice, since statistical subtraction is only performed in the absence of spectroscopic redshifts, hence these are the results we report.  The latter method accounts for the redshift incompleteness of the spectroscopic LF.  Since  incompleteness has been accounted for in the {\sc GalaxyCount} derived model we also tested  the latter technique.  The results were not significantly different since the magnitude limit of the 2dF input catalogue is bright enough that observations are not much affected by the magnitude dependent redshift incompleteness. 

\begin{figure}
\centering \includegraphics[scale=0.4,angle=0]{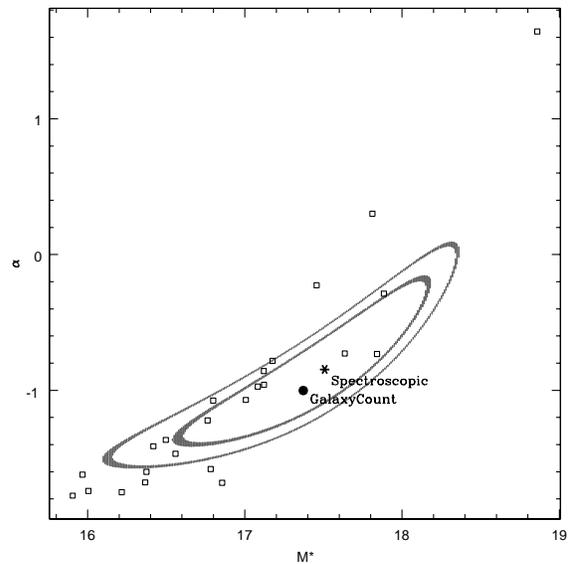}
\caption{The best fitting Schechter functions as described in the text.  The contours show the 68 and 90 per cent confidence limits of the spectroscopic LF.}
\label{fig:lf}
\end{figure}

The LFs were computed following the method described in \citet{ell04}.  \citet{sch76} functions were fit for all galaxies $b_{{\rm j}} \le 19.4384$, i.e.\ the magnitude limit of the 2dFGRS at this position, through minimisation of  \citet{cas79} statistic.  The resulting fits to the characteristic magnitude, $M^{*}$, and the faint end slope, $\alpha$, are shown in Figure~\ref{fig:lf}.  The {\sc GalaxyCount} derived Schechter function is inside the 68 per cent confidence limit of the best fitting spectroscopic results.  Only one offset field is far inside the 68 per cent confidence ellipse, with another 5 offset fields just inside, and another 18 further out.  Thus we expect {\sc GalaxyCount} to be more reliable than an offset field $\sim$75 per cent of the time.  Forty per cent of the offset fields lie outside the 90 per cent confidence limits.   Note that in general it is very difficult to get time allocation committees to grant more than a single offset field per cluster, and there is no way of knowing \emph{a posteriori} whether a `good' offset field has been observed.

\begin{figure}
\centering \includegraphics[scale=0.3,angle=270]{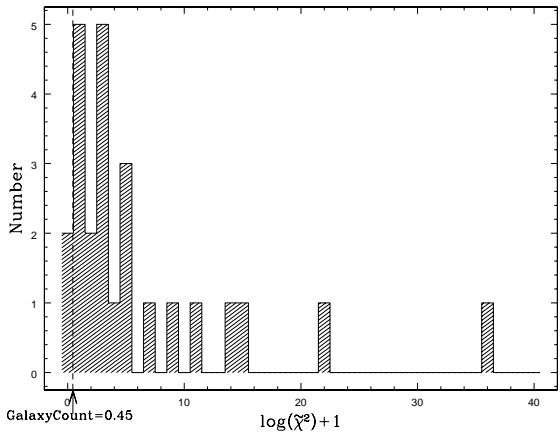}

\caption{Histogram of the $\tilde{\chi}^{2}$ difference between the statistically subtracted LFs and the spectroscopic LF.  The value for {\sc GalaxyCount} is overlaid and is in the upper 25 per cent of the offset fields.}
\label{fig:chi}
\end{figure}

Another way of comparing the results is calculate the reduced $\chi^{2}$ difference between the statistically subtracted LFs and the spectroscopic LF.   These results are presented in Figure~\ref{fig:chi}, which shows a histogram of the $\tilde{\chi}^{2}$ for the 24 offset fields, with the value for {\sc GalaxyCount} overlaid.  {\sc GalaxyCount} is placed in the upper 25 per cent of the offset fields, and occurs near the peak of the distribution, i.e.\ it accurately reflects the expected average result that would be obtained from subtracting a large number of offset fields.

Thus {\sc GalaxyCount} provides an accurate and efficient way to estimate the contribution of foreground and background galaxies to cluster LFs.  In many cases expensive spectroscopic follow-up or off-source background observations may be by-passed and replaced with background estimates from {\sc GalaxyCount} combined with an accurate determination of the completeness of the observations.

\subsection{Comparison to the Hubble deep fields}
\label{sec:hdf}

\begin{figure}
\centering \includegraphics[scale=0.4]{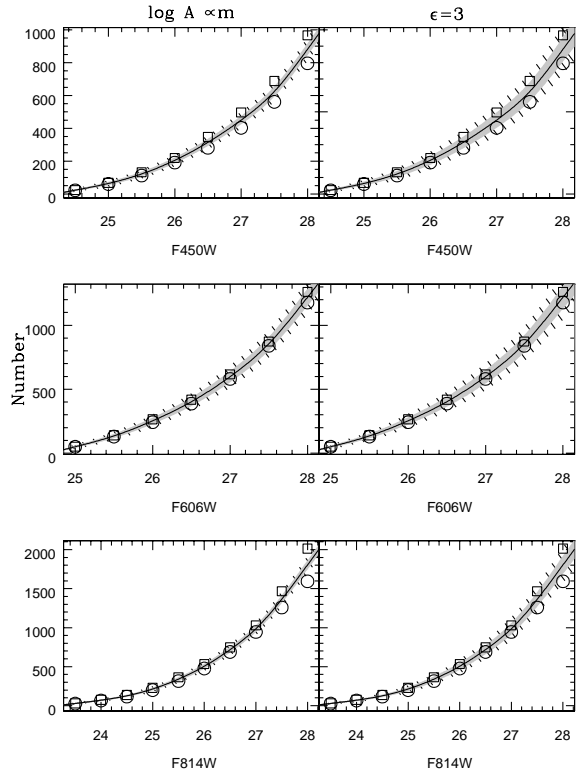}
\caption{The difference between the Hubble deep fields N (squares) and S (circles), compared to the expected standard deviation (1$\sigma$=grey shaded area, 2$\sigma$=hatched area).}
\label{fig:hdf}
\end{figure}

The Hubble Deep Field North (HDF-N) has higher galaxy number counts than the Hubble Deep Field South  (HDF-S, \citealt{met01}).  Figure~\ref{fig:hdf} compares the difference between the two fields to the expected standard deviation as predicted by {\sc GalaxyCount}, with the number counts matched to the average of HDF-N and S.  The two fields are within 1$\sigma$ of each other when using the $\epsilon=3$ model.  Thus it seems that cosmic variance is enough to account for the difference between the two fields.  {\sc GalaxyCount} will be updated to include counts from the Hubble Ultra Deep Field when these become available.

\section{Discussion}
\label{sec:discuss}

\subsection{General results}

We have presented a model to predict the galaxy counts, variances and completeness of observations as a function of magnitude, waveband and area.  The models have been compared to observed number counts and densities calculated using the SDF.  It was found that the model will accurately predict the standard deviation on a given number of galaxies.  Hence for any application for which the number of galaxies is known \emph{a priori} the calculator should accurately determine the appropriate cosmic variance.  

For observations (such as star counts) for which the number of galaxies is not known, more caution must be exercised in using the models.  The derived standard deviation is strongly dependent on the number of galaxies.  However at faint magnitudes an accurate estimate of the number counts can be difficult to determine since the scatter in values published in the literature varies more than the predicted cosmic variance (for example in the $B$ band).  The systematic errors introduced into the number counts are propagated into the resulting standard deviation.  In practice this means that the statistical subtraction of galaxy counts may be dominated by systematic uncertainties rather than cosmic variance in the $B$ band.  However,  for brighter magnitudes where completeness is not an issue the predicted number counts are reliable.

The problem may be alleviated with future, more precise determinations of the faint end number counts.  Until such data is available it is up to the user to choose wisely the number counts which he or she wishes to use and to be aware of the possible systematic uncertainties introduced at faint magnitudes.

The $\epsilon=3$ model was found to predict the standard deviation accurately.    This model has a monotonic decrease in log $A$ with magnitude at bright magnitudes, but flattens off at faint magnitudes.  Such a trend has previously been observed in several studies (e.g.\ \citealt{bra98}; \citealt{pos98}), although many other studies find no such trend (e.g.\ \citealt{mcc01}; \citealt{wil03}; \citealt{coil04}).   Figure~\ref{fig:angcorrmodel}  shows that for the data compiled from the literature a simple monotonic model seems to be a better fit.  However, we are extrapolating $A$ to fainter magnitudes than the measurements, thus it may be that some flattening is required at very faint magnitudes, beyond the limit of the data.  Note that the counts at faint magnitudes dominate the statistics, thus we are testing the currently unknown functional form of the relation at these magnitudes.

An $\epsilon=3$ model implies a strong evolution in the clustering of galaxies since $z=1$.  However, we caution that the physical interpretation of the model should not be taken too literally, since the models do not fit $A$ at all magnitudes, rather the model was used allow a flattening of $A$ at the faint end.  Other studies have reported that a high value of $\epsilon$ is required to fit the faint end of the $A$ scaling relation (\citealt{mcc00b}; \citealt{mcc01}; \citealt{coil04}), but note that these  models cannot fit both the bright and the faint ends of the relation simultaneously.  A precise determination of the value of $A$ at faint magnitudes awaits future measurements.  We remind the reader, however, that the models used in {\sc GalaxyCount} can accurately reproduce the measured standard deviation of galaxy counts in the SDF, despite the uncertainty in the precise functional form of $A$.

{\sc GalaxyCount} has a demonstrated practical use as shown by our reanalyses of the star counts in NGC\,300, and the LF of Abell 2734.  The calculator will prove useful for any observations subject to an unknown background or foreground of galaxies, and in many cases should obviate the need for expensive `blank-sky' observations.





\subsection{When is cosmic variance significant?}

\begin{figure}
\centering \includegraphics[scale=0.4,angle=0]{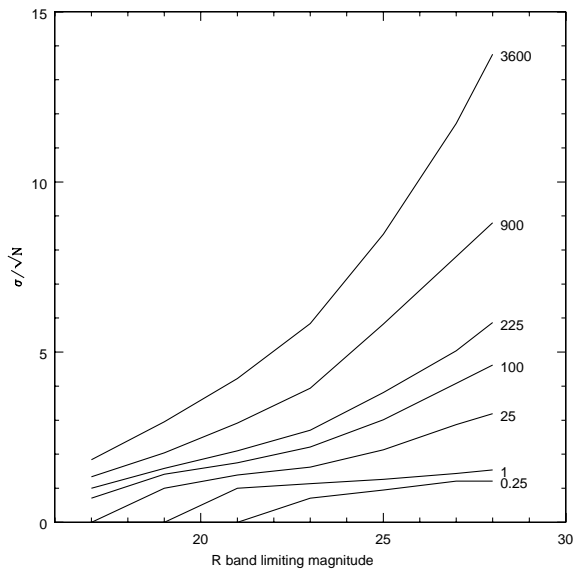}
\caption{The factor by which the total standard deviation dominates over the Poissonian error, $\sigma/\sqrt{N}$, as a function of $R$ band limiting magnitude for various field sizes (labelled in the plot in units of square arcminutes).}
\label{fig:cosvar}
\end{figure}

An important consideration is when does cosmic variance need to be accounted for?  In some circumstances the Poissonian component of the variance will dominate over cosmic variance due to clustering, whereas in others the excess uncertainty will be very important.  This has been addressed throughout the paper by presenting $\sigma/\sqrt{N}$, the factor by which the total standard deviation is greater than the Poissonian standard deviation.  This factor increases with the area and depth of the survey (see equation~\ref{eqn:var}).  We show the form of the dependence in Figure~\ref{fig:cosvar} for the $R$ band predictions as calculated in Table~\ref{tab:main}. 

The non-Possonian contribution quickly becomes important as the surveyed area or the depth of the survey increases; the exact point at which it becomes non-negligible depends on the combination of these variables and the accuracy needed.  For example for a survey reaching a limiting magnitude of $R \le 26$ mag, Poissonian variance will be within a factor of 2 of the true variance only if the area is  $\lsim 10$ arcmin$^{2}$.  Other conditions can easily be tested using the values given in  Table~\ref{tab:main}.  

\subsection{Future work}

{\sc GalaxyCount} will be kept up to date with the inclusion of the latest number densities and angular correlation functions as they are published, including the Hubble Ultra Deep Field number counts.   In addition it will be extended to include predictions for specific targets such as Ly-$\alpha$ sources at high $z$, specific observing conditions such as HST/ JWST observations with a fixed point spread function, and ground-based AO-corrected observations, etc.  An extension is also planned to include predictions of star-counts as a function of co-ordinates, magnitude and FOV.  All future upgrades will be available from the web-site.  

\section*{Acknowledgments}
We thank the referee for very useful comments which have significantly improved this paper.  We acknowledge the very useful Durham Galaxy Counts web pages, data from which were used extensively in this paper.  We thank Nigel Metcalfe for maintaining this useful resource.  We are grateful to Stefano Andreon for useful comments on an earlier draft of this work.  We thank Jon Squire for making his {\sc java} cubic spline fitting code freely available.   SCE acknowledges PPARC funding, partially through the AAO fellowship and partially through grant PP/D002494/1 ÒRevolutionary Technologies for Optical-IR AstronomyÓ.

\bibliographystyle{scemnras}
\bibliography{clusters}

\appendix

\section{Sources of data for number counts and angular correlation functions}
\label{app:sources}

\subsection{Number density}
\label{app:nsources}

The expected number counts of galaxies in the $UBRIK$ bands has been calculated
using a large number of published surveys of galaxy number density, which have been compiled from the following papers:
$U$ band, \citet{koo86}, \citet{guh90}, \citet{madd90}, \citet{son90}, \citet{jon91}, \citet{arnou01}, \citet{met01},   \citet{cap04}; 
$B$ band, \citet{kro78}, \citet{jarv81}, \citet{koo86}, \citet{pet86}, \citet{stev86}, \citet{tys88}, \citet{hey89}, \citet{jon91}, \citet{lil91}, \citet{met91}, \citet{met95}, \citet{ber97}, \citet{met98}, \citet{arnou99}, \citet{arnou01}, \citet{hua01}, \citet{kum01}, \citet{met01}, \citet{yas01}, \citet{lis03}, \citet{cap04}, as well as the Revised Shapley Ames and the Zwicky cat. both from the Durham galaxy counts web pages; 
$R$ band, \citet{cou84}, \citet{hal84}, \citet{inf86}, \citet{koo86}, \citet{stev86}, \citet{yee87}, \citet{tys88}, \citet{jon91}, \citet{met91}, \citet{pic91}, \citet{cou93}, \citet{steid93}, \citet{dri94}, \citet{met95}, \citet{sma95}, \citet{ber97}, \citet{hog97}, \citet{met98}, \citet{arnou99}, \citet{arnou01}, \citet{hua01}, \citet{kum01}, \citet{met01},  \citet{cap04};
$I$ band, \citet{hal84}, \citet{koo86}, \citet{tys88}, \citet{lil91}, \citet{dri94}, \citet{caser95}, \citet{dri95}, \citet{glaz95}, \citet{lef95}, \citet{sma95}, \citet{mam98},  \citet{pos98}, \citet{arnou99}, \citet{arnou01}, \citet{met01}, \citet{yas01}, \citet{cap04};
$K$ band, \citet{mob86}, \citet{gard93}, \citet{soi94}, \citet{djo95}, \citet{glaz95}, \citet{mcl95}, \citet{gard96}, \citet{hua97}, \citet{mou97}, \citet{sar97},  \citet{ber98}, \citet{szo98}, \citet{mam98}, \citet{mcc00}, \citet{vai00}, \citet{mar01}, \citet{hua01}, \citet{koc01}, \citet{kum01}, \citet{mai01}, \citet{met01},  \citet{sar01}, Vandame et al.\ (2001, preprint, astro-ph/0102300), \citet{min05}, and the 2MASS $K$s counts from the Durham galaxy counts web pages.
The majority of the above number counts were retrieved from the Durham galaxy counts web pages\footnote{http://star-www.dur.ac.uk/~nm/pubhtml/counts/counts.html}.

\subsection{Amplitude of the angular correlation function}
\label{app:a}

Measurements of the angular correlation function were taken from: $U$ band, \citet{efs91}; $B$ band, \citet{koo84}, \citet{efs91}, \citet{roc93}; $R$ band, \citet{efs91}, \citet{roc93}, \citet{bra95}, \citet{hud96}, \citet{vil97}, \citet{woo97}, \citet{roc99a}, \citet{con02}; $I$ band, \citet{efs91}, \citet{pos98}, \citet{cab00}, \citet{wil03}, \citet{coil04}; $K$ band, \citet{bau96b}, \citet{car97}, \citet{roc99b}, \citet{kum00}, \citet{mal05}.

\end{document}